\documentclass[12pt]{article}

\usepackage{sbc-template}
\usepackage{graphicx,url}
\usepackage[utf8]{inputenc}

\title{Musical Smart City:\\Perspectives on Ubiquitous Sonification}

\author{Pedro Sarmento\inst{1}, Ove Holmqvist\inst{2}, Mathieu Barthet\inst{1}}

\address{Centre for Digital Music -- Queen Mary University of London
\nextinstitute
  Holonic Systems Oy.
  \email{\{p.p.sarmento, m.barthet\}@qmul.ac.uk, ove@holonic.systems}
}

\begin{document} 

\maketitle

\begin{abstract}
Smart cities are urban areas with sensor networks that collect data used towards efficient management. As a source of ubiquitous data, smart city initiatives present opportunities to enhance inhabitants' urban awareness. 
However, making sense of smart city data is challenging and there is a gap between available data and end-user applications. Sonification emerges as a promising method for the interpretation of smart city data and the production of novel musical experiences. In this paper, we first present the smart city paradigm. We then cover the topics of ubiquitous and mobile music, followed by an overview of sonification research. Finally, we propose an approach entitled ubiquitous sonification and present the initial design of a speculative use case for musical smart city systems, leveraging user and urban data to inform behaviour.
\end{abstract}

%---------------------------------------------------
\section{Introduction}\label{Introduction}
The rapid increase in the availability of large volumes of data justifies the search for new methods to facilitate its interpretation. In a smart city environment, data-gathering devices can range from user products to infrastructure and service sensors, thus congregating multiple modalities. In the context of the internet of things, data visualisation techniques for smart city data have proliferated \cite{Ramos2018, Jing2019}, but much less work has been done to leverage the audio modality.  Due to the density and heterogeneity of the data involved, there is a need for new means of representation in order to support (or extend) existing visualization techniques, allowing users to \textit{get away from the screen} and preserve (or even increase) awareness. Sonification techniques \cite{Kramer1999} can be considered well-suited for this task, as they provide an alternative to the visual domain. Furthermore, the smart city paradigm involves a source of ubiquitous data, present everywhere. This notion also takes into account the fact that its inhabitants' data is an intrinsic part of the system.
%thus implying that users represent a source of data. 
This idea of ubiquitousness of data shares conceptual common points with the field of ubiquitous music, described as the use of musical environments that integrate different types of users and objects \cite{Pimenta2009}. Ubiquitous music argues that in modern societies music is available everywhere, and that, from a musical perspective, there is no possible dissociation between agents and devices, both contributing equally to the final outcome. Thus, a combination of sonification and ubiquitous music techniques emerges as a possible solution towards the representation of data in the smart city context. It is worth clarifying that this work differs from environmental soundscape studies within the city \cite{Steele2019} and from urban sound monitoring \cite{Bello2018}. Indeed, we propose to generate new sound content based on urban related data, and do not focus on analyzing sounds from the city.
This paper begins with Section \ref{Smart City}, presenting a broad description of the smart city archetype. In Section \ref{Ubiquitous Music}, the topic of ubiquitous music is addressed. Section \ref{To Sonify or To Musify?} starts with an overview of sonification research, presenting its definitions and taxonomies. Moreover, it outlines interactive sonification and musification as subsets of sonification, and provides a comparison between the approaches in musification and sonification. In Section \ref{Musical Smart City} we propose the novel concept of ubiquitous sonification which integrates principles from sonification, and ubiquitous and mobile music. We then describe a musical smart city system that relies on machine learning techniques for data dimensionality reduction or cross-modal mapping between data and sound. Furthermore, Section \ref{Discussion} discusses the aforementioned topics as well as the opportunities, challenges and paths for ubiquitous sonification. Finally, Section \ref{Conclusions} concludes with emphasis on the proposed approach, presenting future research directions.

%-----------------------------------------
\section{Smart City} \label{Smart City}
Dainow states that there is not an unified, formal definition of what constitutes a smart city \cite{Dainow2017}, while some authors even argue that a definition is impossible \cite{Albino2015}. Nonetheless, smart cities are often characterized by the presence of a ubiquitous and heterogeneous network of sensors, which provides information about inhabitants and their environment. Smart city sensors are `\textit{embedded in the civic environment, worn on the person and implanted within the body}' \cite{Dainow2017}. Such urban areas encompass wireless sensor networks that collect multiple types of data, which is used in the process of governance, decision-making and planning. Broadly, a smart city ecosystem may consist of sensors for infrastructures, transportation/mobility, environment, services  and user devices (such as smart phones, smart watches and smart home appliances) \cite{Anthopoulos2017}. This system is underpinned by computational power sufficient to process the data acquired. Recently, a growing number of studies point towards the profusion of smart city initiatives across distinct geographical locations as being responsible for shaping new and already existing urban settlements \cite{Cugurullo2018}.
\section{Ubiquitous and Mobile Music}\label{Ubiquitous Music}
Supported by technological advances regarding music generation, distribution and consumption, the field of ubiquitous music is supported by the notion that in modern society music is available everywhere \cite{Holmquist2005}.
%The emergent field of ubiquitous music is characterised by comprising several areas such as sound and music computing, human computer interfaces, creativity studies and music education \cite{Keller2019}.\\
By definition, ubiquitous music is comprised of musical computational environments that allow the integration of multiple users, devices, sound sources and activities \cite{Pimenta2009}. It stands at the `\textit{intersection of mobile and networked music with ubiquitous computing technology and concepts}' \cite{Weiser1991}, where technology, although not visible to the user, is embedded in everyday objects \cite{Mandanici2019a}. As stated by Pimenta et al., in ubiquitous music `\textit{a device is not a passive object that a musician can play, but an agent in a dynamical system that adapts itself to the musical activity, to the local environment and to other agents that interact with it}' \cite{Pimenta2009}. This contrasts with the most commonly adopted vision within computer music research, whereby the musical instrument is seen as the ideal `\textit{metaphor of interaction}'. %By considering a device as an agent within a musical system, on the same level as a user, an analogy can be made here with the notion of actor-network theory in the context of smart cities, presented in the previous section, whereby inhabitants and digital devices also hold equal status.

%Furthermore, this notion also shares multiple synergies with the exposed concept of autopoietic systems with respect to smart cities, where communicative events function as the fundamental pattern that characterizes societies, constantly being generated and regenerated. 
Of close relation to ubiquitous music is the internet of musical things, described as `\textit{networks of musical things (e.g. smart musical instruments or wearables) that support the production and/or reception of musical content}', focused on the interactions between audiences and musicians \cite{Turchet2018}. Strong links are also shared between ubiquitous music and the field of mobile music, the latter focused on the development and usage of mobile devices that act as interfaces for music creation and listening \cite{Essl2017, Gaye2006}. As demonstrated by Bryan et al. with MoMU, a mobile music toolkit implemented for iPhone's OS, these devices present opportunities for new means of musical expression \cite{Bryan2010}, by leveraging mobile sensor data (e.g. accelerometers, compasses and location tracking technology). A particular asset of mobile music is its location-based nature. In this sense, location-based music can be described as sound experiences, normally GPS-driven, in which users' movement in specific zones triggers the playback of sound or music \cite{Hazzard2015}. In their project of a mobile musical soundtrack, Hazzard et al. demonstrate how locative technologies can be used to create rich interactive musical experiences, suggesting that this use case could be transferred to other settings such as a daily commute in a city environment \cite{Hazzard2015}.

%This notion is relevant when framed in the context of smart cities where inhabitants location is a crucial information for the functioning of the whole system (e.g. traffic lights controlled by the amount of inhabitants in vehicles and pedestrians in a given location). 

%In ubiquitous music, musical activity can be seen as the characteristic pattern of the system, and a musical action from one agent/device constantly generates a musical response from other agent/device in a self-sustaining cycle.\\
%In order to conceptually address the notion of  

% DATA TO SOUND OR DATA TO MUSIC
\section{To Sonify or to Musify?}\label{To Sonify or To Musify?}

%Historically, there are many examples that show that sound was the first choice for human beings for the representation and communication of information. As stated by Worrall, `the idea that sound can convey information predates the modern era, and certainly the computational present' \cite{Worrall2018}.\\
%Auditory display is defined as `the use of sound to display quantitative information', being closely related to the concept of sonification. On this topic, Frysinger  argues that, whereas visual displays had long been supported by a strong psychophysical framework, the auditory counterpart is still considered somewhat less intuitive \cite{Frysinger2005}.\\

\subsection{Sonification}\label{Sonification}
The emergence of the research field of sonification is marked by the occurrence of the first conference of the International Community for Auditory Display (ICAD) in 1992 held in Santa Fe, USA and founded by Gregory Kramer \cite{Kramer1999}. 
Since its beginnings, sonification has seen a rise in popularity \cite{Supper2012}, linked to an increase in the availability of \textit{big data} and the consequent notion that humans expect additional ways to enhance the comprehension of their surroundings \cite{Scaletti1991}.\\
The most popular definition of sonification can be posed as `\textit{the use of non-speech audio to convey information}' \cite{Kramer1999}.
According to \cite{Hermann2011}, different types of sonification techniques can be classified as: (i) audification, where data is mapped to sound pressure levels, thus becoming an audio waveform \cite{Dombois2011}); (ii) parameter-mapping sonification, in which each of the data points are mapped to parameters of a sound event, being considered the most common technique of sonification  \cite{Grond2011}; (iii) auditory icons, understood as aural metaphors in which the sound that is heard is a representation of an event, thus informing the listener of its occurrence, assuming prior knowledge about the link between sound and event \cite{Brazil2011}, 
% . This encompasses earcons, where there is no knowledge about the metaphor \cite{McGookin2011})
and (iv) model-based sonification, the use of an acoustic model that generates an output when excited, comprising a set of instructions towards interaction \cite{Hermann2011b}.

Other authors present different subdivisions of the field \cite{Barrass2012}, referring techniques such as sinification (mapping data to sine-tones), MIDIfication (data mapped to MIDI notes), stream-based (granular synthesis techniques for data mapping), vocalization (use of synthesized vowel sounds), iconification (utilizes auditory metaphorical connotations), although it can be argued that these techniques are subsets within parameter-mapping sonification, and iconification can be seen as a class of auditory icons \cite{BonetFilella2019}.

\subsection{Interactive Sonification}\label{Interactive Sonification}
Interactive sonification can be described as a specialized research topic within the field of sonification, in which a human user modifies the sonification process in an interactive control loop  \cite{Yang2019}.
According to Hermann and Hunt, interactive sonification is defined as `\textit{the discipline of
data exploration by interactively manipulating the data’s transformation into sound}' \cite{Hermann2004}. 
%As a research agenda, the authors list six major fields that are considered worth exploring: \textit{Interactive Perception} (how humans deal with different modalities and how does a user's activity influences perception), \textit{Multi-modal Interaction} (`how information should be distributed to different modalities in order to achieve the best usability'), \textit{Interactive Sonification System Analysis} (how the efficiency of `sensor data acquisition and processing, real-time computation of data transformations and rendering of sonifications' could be maximise), \textit{User Learning} (how user engagement and consequent user learning should be measured), \textit{Evaluation} (concerning evaluation methods for the system) and \textit{Ideas and Applications} (exploring new projects with `the potential of bringing computing to a new level of naturalness and depth of experience for the user'). Also, the authors argue that an `interactive sonification system is a special kind of virtual musical instrument, one in which its properties and behaviour depend on the data under investigation, being primarily played with the purpose of learning more about the data' \cite{Hermann2005}.\\
Of great relevance to the field is the Interactive Sonification Workshop. A survey of papers published in its recent editions suggests an emphasis on a more practical, information-driven approach of interactive sonification, comprising works mostly concerned with health issues and biofeedback (e.g. Parkinson and tremor diseases \cite{Schedel2016}, ECG and heart conditions \cite{AldanaBlaco2019}, blindness \cite{Radecki2016}) and mobility (e.g air traffic control \cite{Ronnberg2016}).

\subsection{Musification}\label{Musification}
Barrass defines musification as a sonification technique that uses scales, chords, key and tempo changes \cite{Barrass2012}, however it can be argued that this definition doesn't account for all compositional practices (e.g. musique concrète). In her doctoral dissertation, \textit{Data Sonification in Creative Practice}, Bonet follows an approach on sonification from a more artistic perspective, extending the notion of musification as a sonification that is subject to musical constraints \cite{BonetFilella2019}. Within this scope, a few possible interpretations can be considered: one that comprises a purely functional sonification that is bound to musical principles, other that is solely focused on artistic purposes, and something that an approach in-between. Bonet supports her argument with Varèse's definition of music as `organized sound' \cite{Risset2004}, thus implying that a `\textit{musification is an organised sonification}'. %Furthermore, the author presents some examples of relevant works within this spectrum that use as input data sources such as natural organisms \cite{Takahashi2007}, the human body \cite{Tanaka2012}, and weather and climate \cite{Giannachi2012}.
%, the environment \cite{Abenavoli2012}, and astronomy and particle physics \cite{Cherston2016}.\\ 
%
A framework for the composition of musifications, entitled \textit{Data-Mapping-Language-Emotion}, is described by the author, `crafted to suit the specific requirements of composers working with a scientific method such as sonification' \cite{BonetFilella2019}. Within the framework's first step, \textit{Data}, Bonet stresses the need for a comprehensive understanding of the data by the sound designer/composer, concluding that not all types of data are equally suitable for sonification. Concerning \textit{Mappings}, the process whereby data is transformed into audio, thus becoming perceivable, the author highlights that this is a core stage, often incorrectly understood as the whole sonification process, in which knowledge about human auditory perception and psychoacoustics is important. Due to the enormous range of mapping possibilities, this is also considered the most creative aspect of the technique. In \textit{Language}, the selection of the musical language chosen to transmit the data is addressed, whereby the first must serve the latter. The author claims that `\textit{the aesthetics of the sonification should be appropriate to the purpose of sonification; alarms should be disruptive but displays for long-term monitoring should not be irritating}'. Regarding \textit{Emotion}, Bonet reflects about the parameters that contribute to the storytelling of the sonification.
%\textit{Blyth-Eastbourne-Wembury}, \textit{The Sonification of Dark Matter}, \textit{The Voice of the Sea} and \textit{Wasgiischwashäsch}, using different approaches in each step of the procedure.

\subsection{From Sonification to Musification}\label{From Sonification to Musification}
On their systematic review of mapping strategies for sonification of physical quantities \cite{Dubus2013}, Dubus and Bresin reflect on the prevalent duality within sonification. Some researchers are concerned with the need of having a stricter definition \cite{Hermann2008}, somehow prioritizing the conveyance of information within the method,  while others are `\textit{willing to step over the border to data-driven music}', supporting a more inclusive perspective and highlighting the importance of musical aesthetics. As described by Walker and Kramer, such divergences are based on the interdisciplinarity of the field, meaning that the sonification process usually involves concepts from both the arts, science and engineering \cite{Walker2004}. Collaborations between researchers and composers (or experts in sound related fields) are frequent within the practice of sonification. Likewise, the increasing significance attached to artistic works in the ICAD programme is also a consequence of this ambivalence \cite{Dubus2013}. Neuhoff noted that `\textit{there are 1,103 conference papers in the ICAD proceedings, from the years 1994 to 2018, and that the word `music' appears in 74\% of these works}' \cite{Neuhoff2019}. In an attempt to find a new definition for sonification, Hermann states that a technique that produces sound signals from data, may be called sonification if and only if \cite{Hermann2008}: the sound reflects \textit{objective} properties or relations in the input data, the transformation is \textit{systematic} (meaning that there is a precise relationship on how the data and possible interactions cause the sound to change), the sonification is \textit{reproducible} and the system can intentionally be used with \textit{different data}, and also be used in repetition with the same data. According to Supper, this definition represents an attempt to `\textit{narrow down the boundaries of the field}', implying that it emphasizes more the scientific aspect of sonification, omitting consideration for its artistic side \cite{Supper2012a}. Another perspective presented by Barrass is concerned with how sonification can be used in design research, stirring sonification \textit{beyond the science laboratory and artistic exhibition}, to create novel products and further investigate human behaviour through interaction with them \cite{Barrass2018}.

\section{Musical Smart City}\label{Musical Smart City}
\subsection{Ubiquitous Sonification}\label{Ubiquitous Sonification}
The combination of topics covered in this article points towards an approach that combines sonification and ubiquitous and mobile music techniques. Thus, the notion of a ubiquitous sonification technique can be posed as a \textit{type of sonification that leverages ubiquitous computing environments}. Following Satyanarayanan's pervasive computing vision, this assumes that a user is permanently immersed in a personal, unobtrusive, digital space that mediates interactions with other surrounding ubiquitous computing devices \cite{Satyanarayanan2001}. It is worth to clarify that a major distinction between ubiquitous music and ubiquitous sonification is that the latter involves a component of information conveyance, aiming to raise awareness in the user, not focusing solely in the musical value of the system.
Furthermore, due to the role of the user in this scenario, the subfield of interactive sonification presents multiple synergies with ubiquitous sonification, the latter following some of the suggestions proposed by Hermann and Hunt \cite{Hermann2004}, namely those concerned with the way humans deal with different modalities and how a user's activity influences perception. Finally, previous works have employed the term \textit{ubiquitous sonification} \cite{Nees2018, Beilharz2009, Macdonald2014}, but they all use the word \textit{ubiquitous} in order to refer to a well-established and widely used sonification process (e.g. the Geiger counter, which aurally displays information about ionizing radiation).  However, our use of the term ubiquitous is different and focuses on the use of data provided from ubiquitous devices.
%This approach is inspired by the work of Tünnermann et al. and the concept of blended sonification, the process of manipulating physical interaction sounds or environmental sounds in such a way that the resulting sound signal carries additional information of interest \cite{Tunnermann2013}. One of the major differences between this technique and ubiquitous sonification comprise the fact that the latter would not be site specific, exploring the smart city exemplar of `data available everywhere', thus the choice of \textit{ubiquitous} as a suitable term.
%On the other hand, the guideline of \textit{calmness} is understood as an important parameter in order to make an ubiquitous sonification system non-irritating.

\subsection{Musical Smart City System}\label{Musical Smart City System}
\subsubsection{Vision}\label{Vision}
 The smart city paradigm, due to the large volume of data generated by its ubiquitous, heterogeneous sensor network, presents a promising opportunity for the application of ubiquitous sonification.  Following the ubiquitous computing proposal, whereby technology is embedded in everyday objects, ubiquitous sonification presupposes that technology enabling sonic interaction is readily available to the user (e.g. through usage of smart devices such as watches, bracelets, phones). Beyond standard daily human interactions and behaviour within a social context, it can be envisioned that in the smart cities of the future, inhabitants will implicitly communicate with smart devices through a digital medium. This relation is also regulated by the inputs received from those same devices and applications.
 %\todo{REMOVE?} One of the major conceptual motivations for ubiquitous sonification is the definition of a smart city as an integrated domain presented in Section \ref{Smart City}. The depiction of the smart city paradigm as an urban digital environment where human agents are positioned on the same ontological level as the rest of (digital) devices, positions smart cities as a very interesting use case for a technique such as ubiquitous sonification.
 
\begin{figure}[h!]
	\centering
		\includegraphics[width=0.7\columnwidth]{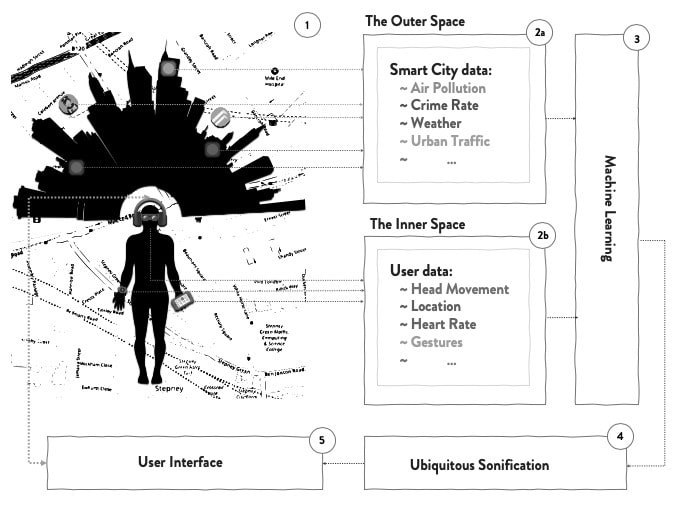}
	\caption{Conceptual diagram of a musical smart city system.}
	\label{fig:BlockDiagram1}
\end{figure}

 A conceptual diagram of a musical smart city system is presented in Figure \ref{fig:BlockDiagram1},
describing a location-based system that is able to increase inhabitants' urban awareness within a city, while accounting for an exploration of their environment in a musical way. Conceptually, its structure comprises \textit{outer space} and \textit{inner space} (respectively, 2a and 2b in Figure \ref{fig:BlockDiagram1}), the first being related with data from inhabitant's surroundings (such as air pollution, crime rate, land usage, urban traffic) and the latter connected to a user's physiological, inertial and location data provided by smart wearable devices and phones (such as heart rate, hand gestures, head movement and location). In order to extract relevant features from large volumes of data, machine learning algorithms are used to achieve dimensionality reduction (depicted as item 3 in Figure \ref{fig:BlockDiagram1}) \cite{Winters2019}. Furthermore, as proposed by Fried and Fiebrink, a cross-modal mapping between data and sounds (e.g. inputs from smart city sector data sets mapped to a large collection of samples, musical motifs or tracks) could be carried out by deep learning algorithms (e.g. deep auto-encoders) \cite{Fried2013a}. 
%The outcome of this process can be seen as a form of sonification.
Processed data is sonified using concepts of ubiquitous sonification (item 4 in Figure \ref{fig:BlockDiagram1}). An initial approach might consider using parameter-mapping sonification for data that concerns the inner space, mapping user data to musical parameters, and model-based sonification for outer space data, creating a musical particle system a user can navigate in. This will also require the development of a framework that accounts for the interactions between inhabitants and environment, relating the musical output of the inner space with the one from the outer space (e.g. user's gestures generating a rhythmic pattern that is fused with the soundscape generated by air pollution data). 
Finally, mediated by a user interface, this will result in a mobile music platform which supports musical mappings from physiological and urban events, raising user's awareness (item 5 in Figure \ref{fig:BlockDiagram1}).

In this setting, what would be a use case for a musical smart city system? A possible scenario envisions a user connected to the musical smart city mobile application while walking in the city. Aurally, this user is able to perceive that the levels of air pollution in her usual route to work are abnormally high and thus decides to take an alternative route, sonically depicted as less polluted by a smoother, sparser soundscape. On her way back, alone at night, the musical smart city application conveys sonic information about which direction to take in order to travel through streets/areas that have lesser indexes of crime rate (e.g. pointing towards the ones covered by CCTV). 

\subsubsection{Challenges}\label{Challenges}
% Highlight the challenges, what kind of data
% Identify IoT data streams and characterize their nature, temporal and spatial scales. Reflect on the usefulness of this for sonification and musification. 
% Identifying user requirements (in any design study we have user requirements). What type of sectors of the city would a user be interest in know more about. 
% Explain why it is important (multimodal data fusion techniques)
% MOVE FROM VISION "Anticipating ..." to challenges
% DATA GATHERING AND 
% DESIGN AND EVALUATION
This vision paper raises many questions and challenges that will stir upcoming research. \textbf{Data gathering and characterisation}: it is important to identify IoT data streams and characterise their nature, temporal and spatial scales, reflecting on their suitability for sonification. \textbf{User requirement identification}: reflect on different sectors of the city environment and the usefulness of sonifying them from a user perspective; take into account what city aspect's users would like to learn more about or experience in an enhanced way using the auditory modality. Of interest are considered topics related with the environment (pollution, waste water treatment, weather and climate) and mobility, but other areas concerning service delivery (health and medical care, crime rate and security) could be posed as relevant research paths. \textbf{AI for sonification/musification}: anticipating large volumes of data from the aforementioned modalities, it is important to select suitable machine learning algorithms to achieve dimensionality reduction. It would also be of interest to study the combination of multiple modalities exploring multimodal data fusion techniques. Unsupervised machine learning techniques could be taken into account in order to explore possible patterns between different types of data. As for sound content creation, mappings between data and music could be mediated by deep learning approaches for music generation, conditioning their output on data. \textbf{Design and evaluation of the system}: the design of user interfaces and suitable protocols for its evaluation should be considered. These methods should investigate the system's ability to convey information about a given subject and also assess its creative and aesthetic merits.
%This vision paper raises many questions that point towards future work directions. First, a thorough selection of the most suitable smart city aspects should be carried out, requiring a comprehensive understanding of the existing sectors. What types of data are most adequate within the context of a musical smart city? Of interest, are topics related with the environment (renewable energy production, pollution, waste water treatment, weather and climate, biodiversity) and mobility, but other areas concerning service delivery (health and medical care, crime rate and security) could be posed as relevant research paths. Furthermore, machine learning techniques that allow a reduced dimension representation of data (e.g. deep auto-encoders, dimensionality reduction) need to be assessed and selected. Moreover, decisions that regard the sonification process would represent an important part of the system. Which techniques, software and tools should be used? What degree of creative freedom should be given to the user? What is the best compromise, if any, between a more information driven approach and the more creative, artistic one? Finally, procedures for evaluation of the system's performance should be formalized. These methods should investigate the system's ability to convey information about a given subject and also assess its creative and aesthetic merits.  The answers to the aforementioned questions pose itself as the main drive for research and future work in this domain.

\subsubsection{Related Work}\label{Related Works}
The project \textit{Sonic City}, by Gaye et al., represents an early exploration of the city environment as an interface for musical expression \cite{Gaye2003}. The authors implemented a wearable prototype that retrieves information about user mobility and maps it to real-time audio processing of urban sounds. Future work points towards the usage of smart devices (e.g. wireless devices with built-in sensors and computational power) instead of the developed prototype, which clearly serves as a motivation for the vision presented in this paper. Furthermore, in the work carried out by Winters et al., the authors present \textit{The Decatur Civic Dashboard}, a multi-modal dashboard for the sonification of data in the context of smart cities using the Web Audio Javascript API as an attempt to turn the process \textit{away from the desktop} \cite{Winters2016}. This approach is supported by a specially built Javascript library called DataToMusic that serves as a helping tool in the sonification process. The generated audio represents an informative complement to the graphical display in the dashboard. From an informative perspective, the work of Winters et al. shares multiple common points with musical smart city, but one of the distinctions could be posed as a difference of focus  on the notion of the city as a musical interface, somehow putting aside the more creative aspect of the process. Moreover, in \textit{The Decatur Civic Dashboard} project, sonification of user data is not emphasized, which in the case of musical smart city represents an important part of the system. Pigrem and Barthet propose the concept of datascaping as the usage of data as a medium in soundscape composition. This technique is used to sonify real-time data of Transport for London API, conveying information about levels of traffic in stations and lines of underground transportation. Interestingly, the authors argue that `\textit{when data used in the production of an artwork describe or correlate with some human activity or state, the people represented by the data hold a participatory role in the realisation of the art work}' \cite{Pigrem2017}. This is an important aspect of musical smart city, whereby data from inhabitants is an integral part of the system. The work in \textit{DataScaping} represents what would be considered a successful case of applying sonification in a specific sector  of a smart city and the authors conclude by suggesting the need for assessment of this techniques in the case of multiple modalities of data. 
Finally, the work of Steele et al. tackles awareness of the role of sound in urban settings \cite{Steele2019}. In \textit{Sounds In The City}, the authors propose two workshops held in Montreal, the first focusing on using sounds to create audible experiences in pedestrian zones, and the second about the preservation of good-quality sound environments. Furthermore, with \textit{SONYC}, Bello et al. presents a system to mitigate noise pollution within the city environment. By employing machine listening techniques over sound recordings collected from \textit{ad hoc} sensors, different outdoor sounds are classified and characterized according to its source, thus supporting further decision making regarding noise mitigation. Despite some common factors, namely the exploration of the sonic dimension within the city context, both projects differ from the work proposed here, which will leverage data (not urban sounds) has source material for the generation of musical content.   

\section{Discussion}\label{Discussion}
In this paper, the technique of ubiquitous sonification was proposed, presenting possible applications that exploit the smart city environment, envisioning a musical smart city system. This system would be user-centered, allowing for the creation of musical content and, at the same time, to raise awareness about inhabitants' surroundings. One of the motivations for this approach would be not only to `\textit{bridge the gap between audiences and artists by blurring the roles of creators and receivers}' \cite{Pigrem2017}, as happens in participatory art, but also to follow an analogy between layman and specialist, concerning data interpretation, by inviting the user towards an auditory manipulation of information.
As discussed in Section \ref{From Sonification to Musification}, in which the dichotomy between a more formal, information-driven approach, and a more artistic, creative-based one in the field of sonification was addressed, some considerations are worth to be discussed, which might benefit the concept of ubiquitous sonification. Following a pure empirical/informative perspective, Bonet's suggests an analogy between Shannon-Weaver's model in \textit{The Mathematical Theory of Communication}, in which the act of communication is composed of an information source, a transmitter, a channel, a receiver and a destination, and the conceptual process of sonification. According to the author, in this scenario, the source of information would represent the sonification designer's output, the transmitter (or encoder) would be equivalent to the chosen mappings, the channel would be considered the used musical language, the receiver (or decoder) would stand as the knowledge of the mappings and musical language utilized, and the receiver would function as the listener \cite{BonetFilella2019, Shannon1949}. Within this analogy, being a mathematical theory of communication, the emphasis is kept on the conveyance of information between the transmitter (the sonification designer) and the receiver (the listener). Concerning a more artistically-driven approach, the focus should be stressed on the channel (the musical language). As McLuhan postulates in his work \textit{Understanding Media: The Extensions of Man}, the global impact generated by the channel, or the media environment, is more significant than the content it conveys \cite{McLuhan1964}. Affirming that `\textit{the medium is the message}', McLuhan was clearly addressing the most popular media of his time, television and radio. The author postulates that the impact of the media itself, the \textit{channel}, is higher than that of single programs or content they emit. An analogy can be made in the case of a music driven ubiquitous sonification, thus inferring that, from a global perspective, the choice of the musical language, the structure of the `creative material', would be the most central aspect of the process. An ubiquitous sonification application, framed within the context of smart cities, could perhaps tackle these questions by allowing the user to navigate between both a more information or artistically-driven output.

\section{Conclusions}\label{Conclusions}
In this vision paper, the paradigm of smart cities was addressed from a data and agent perspective, framing it as an integrated domain of both digital and human actants. We established a conceptual link between the latter and aspects of the fields of ubiquitous and mobile music. Moreover, a literature survey concerning sonification revealed common used techniques and approaches in the field. We addressed the distinctions between the specifications of musification and interactive sonification, encouraging a discussion about the concepts behind those approaches. Furthermore, we defined ubiquitous sonification as a concept integrating approaches in sonification and ubiquitous and mobile music. Finally, a vision towards a musical smart city system was described, pointing towards future research questions and directions.

\section{Acknowledgements}\label{Acknowledgements}
This work is supported by the EPSRC UKRI Centre for Doctoral Training in Artificial Intelligence and Music (Grant no. EP/S022694/1).

\bibliographystyle{sbc}
\bibliography{sbc-template}

\end{document}